# The unification of Pythagorean theorem for electronic orbitals with Kepler's law for planetary orbits

Kunming Xu

Environmental Science Research Center,

Xiamen University, Xiamen 361005,

Fujian Province, China

**Abstract:** In the context of two-dimensional spacetime within a helium atom, both 1s electrons are characterized by wave functions that observe duality equation. They are symmetric, orthogonal and interwoven, forming a dynamic rope structure at any moment. Instead of elliptical orbit of planets around the sun, electronic orbitals take the form of matter state transformation cycle. While the kinematic movement of planets is governed by Kepler's first law, electronic transformation obeys Pythagorean theorem, both being equivalent in physical principle. The atomic spacetime is a continuous medium of electron clouds in synchronized differential and integral processes that are implemented by smooth trigonometry. In order to integrate this new approach with conventional physics, the author translates the pattern of electronic motion in the atomic spacetime into spherical volume undulation in Euclidean geometry and calculates the probability density of an electron within the sphere from the classical perspective. From the primary wave function of a 1s electron, the author also tries to derive the mathematical expression of central force that guides the surrounding bodies along the orbits. The result is exciting and surprising that questions the exactness of the venerable Coulomb's law.

## 1. The rope structure of electrons

It has been proposed that the wave function of electrons within a helium atom obey duality equation of

$$\frac{\partial^2 \Omega}{\partial t^2} = v^2 \frac{\partial^2 \Omega}{\partial l^2}, \tag{1}$$

where $\Omega$ is a wave function, $v$ is a velocity dimension, $t$ and $l$ are time and space dimensions respectively [1-3]. By separating time and space variables, this partial differential equation may be decomposed into two synchronized ordinary differential equations as

$$\frac{d^2 \Psi}{dt^2} = -\omega^2 \Psi, \tag{2}$$

$$\frac{\partial^2 \psi}{\partial l^2} = -\frac{1}{r^2} \psi, \tag{3}$$

$$v = \omega r, \tag{4}$$

where $\Psi$ denotes time component and $\psi$ denotes space component of the electrons, and $\omega$ is angular velocity while $r$ is orbital radius.

Because wave function evolution follows differential and integral operations, there are interesting relationships between a wave function and its first and second derivatives. As shown in Figure 1(a) for equation (2) in two-dimensional spacetime, dimension $d^2\Psi/dt^2$ must be equivalent to dimension $\Psi$, i.e. performing differential operations twice upon a wave function returned it to the original wave function. Hence we derived $\omega^2 = -1$ from the equation formulation so that $\omega$ was a complex number identifier in the wave function. Moreover, since operator $d^2/dt^2$ performing upon $\Psi$ constituted a complete loop in the two-dimensional spacetime, dimensions $\Psi$ and $-d\Psi/dt$ must be symmetric within the looping cycle. It followed that if $\Psi_1$ was a valid solution to equation (2), then $-d\Psi_1/dt$ must be a valid solution as well for it was an intermediate step for wave function $\Psi_1$ to evolve towards $d^2\Psi_1/dt^2$. For example, if the first solution to equation (2) was

$$\Psi_1 = C_1 \cos\alpha, \tag{5}$$

where $C_1$ was a time constant, $\alpha$ was a radian angle related to time component of the electron, then the second solution was its first derivative $-d\Psi_1/dt$:

$$\Psi_2 = -C_1 \omega \sin\alpha, \tag{6}$$

$$-\frac{\partial \alpha}{\partial t} = \omega, \tag{7}$$

where the negative sign indicated contrary alignment of $\alpha$ variable with *t* dimension. Likewise, as shown in Figure 1(b) for equation (3) in two-dimensional calculus spacetime, the solutions to it can be compactly written in a complex function

$$\psi = C_2 (\cos\beta + r\sin\beta), \tag{8}$$

$$\frac{1}{r} = \frac{\partial \beta}{\partial l}, \tag{9}$$

where $C_2$ was a space constant, *r* was a complex number notation, and $\beta$ was radian angle related to space component of the electron.

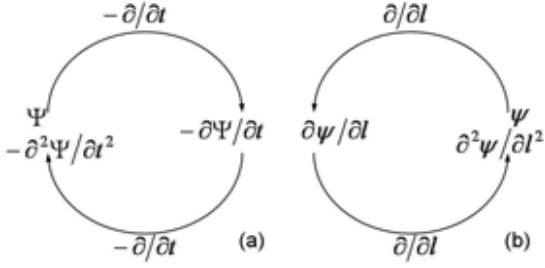

Figure 1. The equivalence between a wave function and its second derivative and the symmetry of a wave function and its first derivative in (a) time component and (b) space component of the electrons within a helium atom.

Combining time and space components altogether, we had four unique roots to duality equation in the following matrix:

$$\begin{pmatrix} \Omega_0 \\ \Omega_1 \\ \Omega_2 \\ \Omega_3 \end{pmatrix} = C_1 C_2 \begin{pmatrix} \cos\alpha \cos\beta \\ -\omega \sin\alpha \cos\beta \\ -v \sin\alpha \sin\beta \\ r \cos\alpha \sin\beta \end{pmatrix}, \tag{10}$$

which had intimate calculus relationships as shown in Figure 2(a). However, if we regarded the differential and integral operations as being carried out simultaneously, then functions $\Omega_1$ and $\Omega_3$ may be cancelled from the list so that $\Omega_0$ and $\Omega_2$ represented both electrons in a helium atom. As shown in Figure 2(b), both wave functions had an intervening spacetime distance of reducing a time dimension while increasing a space dimension, i.e. performing $\int (-\partial/\partial t) dl$ operation upon $\Omega_0$ produced $\Omega_2$ and performing the same operation upon the latter yielded the former, both operations forming a cycle. The cycle was an $\int (-\partial/\partial t) dl$ operator cycle as well as a wave function cycle, or simply a dynamic calculus cycle. Each wave function was the symmetric counterpart of the other in the calculus spacetime. Physically, if we regarded the transformation of $\int (-\partial/\partial t) dl$ upon electrons as a twisting tension, then both electrons formed a rope structure because the calculus operation upon both wave functions was analogous to the common procedure of making a rope such as by twisting a strand of hemp in a fixed direction and then letting it fold back to form two intertwining strands. In this sense, the rope structure was a pithy summary that both 1s electrons were symmetric and intertwined in the calculus spacetime.

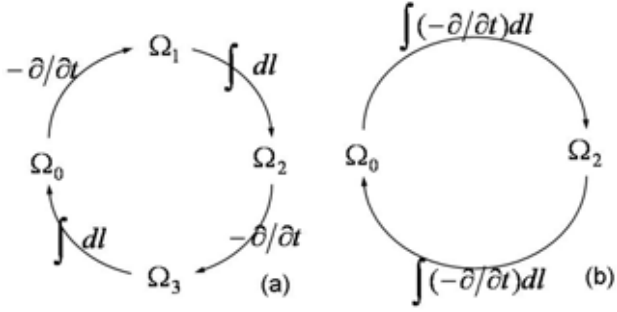

Figure 2. Calculus relationships (a) between four characteristic roots of duality equation and (b) between two representative roots that formed a rope structure.

## 2. Spacetime continuity via dynamic calculus

After defining the wave function of both electrons in a helium atom, we shall further illustrate their motion by examining the course of wave function revolution by dynamic calculus. First of all, let's establish dynamic calculus concept by studying the special relationship between trigonometry and calculus. A typical relation between differential operation and trigonometric function can be expressed by

$$-\frac{\partial \Omega_0}{\partial t} = C_3 \cos\alpha, \text{where } (\alpha \mapsto \frac{\pi}{2} + \alpha), \tag{11}$$

where $C_3$ was a quantity orthogonal to $\cos\alpha$ and the arrow $\mapsto$ indicated the process of radian variable rotation from $\alpha$ to ($\pi/2 + \alpha$). As has been described in the previous report [1], a differential operation is realized through the smooth increment of $\alpha$ angle from $\alpha$ to ($\pi/2 + \alpha$), which of course results in the change of $\cos\alpha$ term into $-\sin\alpha$ term in the end. In that sense, we may express a differential operation in terms of radian angle rotation. Since electronic motion is a dynamic process, here we interpreted the differential operation as a gradual sinusoidal course. By treating $\alpha$ as a continuously changing variable, we may omit the subordinate clause in equation (11) and express a differential operation by a simple trigonometric function, disregarding whether the differential process was carried out completely or not. If it was completely done, then $\alpha$ increased a displacement of $\pi/2$ so that $\cos\alpha$ transformed into $-\sin\alpha$. Because $\alpha$ was a dynamic variable, the difference between both trigonometric terms ($\cos\alpha$ and $-\sin\alpha$) was a matter of $\alpha$ value difference in function $\cos\alpha$. Thus, both $\cos\alpha$ and $-\sin\alpha$ can be used for expressing the differential operation on the left-hand side of equation (11), even though we traditionally associate the final result of the differentiation with $-\sin\alpha$, and the initial condition with $\cos\alpha$. The following two definitions were the dynamic interpretation of calculus:

**Definition 1.** For a trigonometric wave function such as $f(\alpha) = \cos\alpha$ or $f(\alpha) = \sin\alpha$, a differential operation on $f(\alpha)$ with respect to $\alpha$ means increasing $\alpha$ variable a displacement of $\pi/2$ in the function.

**Definition 2.** For a trigonometric wave function such as $g(\beta) = \cos\beta$ or $g(\beta) = \sin\beta$, an integral operation on $g(\beta)$ over $\beta$ means reducing $\beta$ variable a displacement of $\pi/2$ in the function.

Graphically, if we draw wave function $f(\alpha)=\cos\alpha$ in Cartesian coordinates, then increasing $\alpha$ variable a displacement of $\pi/2$ in the function can be interpreted as translationally moving $f(\alpha)$ axis in the positive direction a displacement of $\pi/2$ so that $\cos\alpha$ transforms into $-\sin\alpha$ in the differential operation. Likewise, if we draw the wave function $g(\beta)=\cos\beta$ in Cartesian plane, then reducing $\beta$ variable a displacement of $\pi/2$ in the function can be interpreted as translationally moving $g(\beta)$ axis in the negative direction a displacement of $\pi/2$ so that $\cos\beta$ becomes $\sin\beta$ in the integral operation.

The difference between conventional infinitesimal differentiation concept and the above angle rotation definitions is that the latter traverses the full range of π/2 angle rotation in the trigonometric function whereas the former only captures the terminal state of the trigonometric function at a certain radian value. In this sense, the latter, covering the full course, was actually the dynamic implementation of the former.

According to the dynamic calculus implementation, the derivatives of wave functions of both electrons in helium shell may be expressed as

$$-\frac{\partial \Omega_0}{\partial t} = C_3 \cos\alpha; \tag{12}$$

$$\frac{\partial \Omega_2}{\partial l} = C_4 \sin\beta, \tag{13}$$

where $C_3$ and $C_4$ were orthogonal to the trigonometric functions respectively, and $\alpha$ and $\beta$ were dynamic radian variables related to time and space components of electrons respectively. Because equation

$$-\frac{\partial \Omega_0}{\partial t} = \frac{\partial \Omega_2}{\partial l} \tag{14}$$

always holds, both radian variables must be synchronized and complement satisfying

$$\cos\alpha = \sin\beta, \tag{15}$$

$$\alpha + \beta = \frac{\pi}{2}. \tag{16}$$

These relationships have been illustrated diagrammatically in the previous report [2]. To rehash, we illustrate electronic transformation from wave function $\Omega_0$ to $\Omega_2$ in two synchronized diagrams as shown in Figure 3. For time component, as the electron traveled from point A towards point B along arc ACB, it traversed a time dimension from a dimensionless quantity along X-axis to the angular velocity dimension along Y-axis. At any specific moment of point C on the pathway, the derivative $-\partial\alpha/\partial t$ as expressed by $\omega$ in Figure 3(a) was always orthogonal to radian angle $\alpha$ because they were exactly a time dimension apart. Radian angle $\alpha$ can be expressed by the arc in ACB direction whereas angular velocity was orientated along the radial direction at any moment, hence their orthogonal relationship was that between a circular arc and its radius. Differential operation $-\partial\Omega_0/\partial t$ transformed $C_1\cos\alpha$ into its orthogonal dimension of $-C_1\omega\sin\alpha$ with the complex notation $\omega$ as its dimension identifier; and further differential operation $-\partial\alpha/\partial t$ transformed radian arc $\alpha$ along the circle into its orthogonal dimension of $\omega$ along the radial direction. Both operations were consistent in the usage of $\omega$ as complex number notation that transformed its operand into its orthogonal dimension.

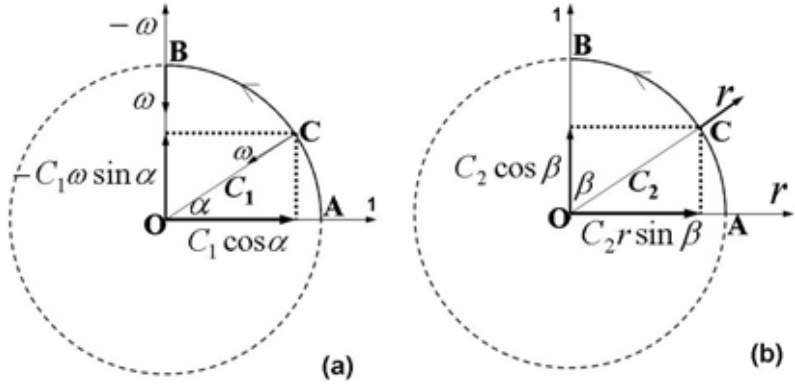

Figure 3. Schematic illustration of the physical significance of (a) angular velocity $\omega$ and (b) orbital radius $r$ in the atomic spacetime.

Likewise, as shown in Figure 3(b), the circular pathway from A to B represented electronic transformation of $\int\Omega_1 dl$ implemented by the continuous decreasing of radian angle $\beta$ from $\pi/2$ to zero in the process. At any specific moment C along the pathway, orbital radius $r$ was orthogonal to arc BC that represented radian angle $\beta$. Integral operation $\int\Omega_1 dl$ transformed $C_2\cos\beta$ into its orthogonal dimension of $C_2 r\sin\beta$ with complex notation $r$ as its dimension identifier; and further integral operation $\int 1/r\,dl = \int d\beta$ established the orthogonal relationship between $r$ and $\beta$ as that between a radius and its circular arc. Therefore, the usage of $r$ as complex number notation was consistent with its derivative expression (9) in calculus. The geometric circle of Figure 3 established electronic motion as a circular one, but it differed from a Cartesian circle in that revolution along the path indicated dimensional change by the rule of dynamic calculus instead of position movement in X-Y coordinates.

During the calculus transformation $\int(-\partial/\partial t)dl$, trigonometric implementation ensured that there was not any break or void or abrupt jump in the course. The atomic spacetime was continuous in that space and time waves were transforming smoothly and continuously from one state to another. Here spacetime continuity had non-classical meaning. To give an analogy, when we said that a father and a son were spacetime continuous, we did not mean that they sat on the same bench and in close touch, but that the son was developed from the seed of the father. Space referred to the biological bodies, not the location or volume of air that they empty off. By the same way, the atomic spacetime continuity did not mean that both electrons occupied two positions close enough, but that both electrons were transforming from each other smoothly without any interruption. Indeed, both electrons occupied the whole atomic spacetime without a clear physical distinction between them. Spacetime continuity referred to smooth calculus transformation implemented by coherent trigonometry. This concept was more fundamental and strict than conventional function continuity at a specific position under Cartesian coordinates.

After introducing dynamic calculus and spacetime continuity concept, we shall make a modification concerning the general solution to duality equation. While four characteristic roots as was shown in equation (10) set the dimensional framework for electrons, every point of C along the circle (Figure 3) constituted a valid solution. Because the pathway of point C was a continuous circle as was determined by dynamic synchronized radian angles $\alpha$ and $\beta$, given a specific pair of complementary $\alpha$ and $\beta$, there was one solution to duality equation corresponding to it. In other words, there were countless solutions to duality equation and the circle constituted the set of all solutions. Among others, two special points along the circle deserved further scrutiny in mathematics.

Table 1. The relationships between space and time dimensions and between both electronic wave functions at the boundary radian values of Figure 3.

| Wave function | Two dimensions | | Electron |
| --- | --- | --- | --- |
| | $\alpha = 0; \beta = \pi/2$ | $\alpha = \pi/2; \beta = 0$ | |
| $C_1 C_2 \cos\alpha \cos\beta$ | $C_1$ | $C_2$ | $\Omega_0$ |
| $-C_1 C_2 v \sin\alpha \sin\beta$ | $C_2$ | $C_1$ | $\Omega_2$ |

As shown in Figure 3 and Table 1, at the initial point A, radian angles $\alpha = 0; \beta = \pi/2$ so that wave function $C_1 C_2 \cos\alpha \cos\beta$ declined into space constant $C_1$ while at the terminal point B, radian angles $\alpha = \pi/2; \beta = 0$ so that the wave function became space constant $C_2$, both constants actually representing both dimensions of time and space respectively. In such calculations, the expression $C_1 C_2 \cos\alpha \cos\beta$ as a wave function should not be construed as conventional multiplication of orthogonal time and space components, instead the multiplication was only a time and space coordinative notation that did not share the common property of a conventional multiplication. For example, when space component $C_2 \cos\beta = 0$, the whole wave function was not equal to zero, but equal to time component $C_1 \cos\alpha$ with space component disappearing away. In addition, at the boundary points of A and B, wave functions only reflected the dimensions of $C_1$ and $C_2$ diametrical constants without involving the identifiers $\omega$, $r$, and $v$ in wave function $-C_1 C_2 v \sin\alpha \sin\beta$ because complex number marker was no longer necessary at those points. It was clear in Table 1 that the difference between both electrons was characterized by $\pi/2$ radian phase, so was that between both space and time dimensions of each electron.

The foregoing discussion demonstrated three ways to express a differential operation. The first was in conventional derivative form such as $-\partial\Omega_0/\partial t$; the second was in dynamic trigonometric term such as $C_3 \cos\alpha$; and the third was in a shorthand such as $\dot{\Omega}_0$ similar to $\omega$ or $r$ expression that connoted complex number notation. These three usages agreed with each other in expressing orthogonal transformation of a physical quantity in the atomic spacetime. By expressing $-\partial\Omega_0/\partial t$ in term of $C_3 \cos\alpha$, we meant: a) that $\cos\alpha$ was a representative trigonometric function whose cosine and sine compositions were determined by the value of radian angle $\alpha$: when $\alpha = 0$, it was a full cosine composition and when $\alpha = \pi/2$, it was a full sine function. This compositional transition was a result of smooth translational movement of Y-axis by a displacement of $\pi/2$ in function $y = C_3 \cos x$; b) that the time component of $\Omega_0$ changed sinusoidally corresponding to the uniform rotation of angle $\alpha$ from 0 to $\pi/2$, i.e. $-\partial\Omega_0/\partial t$ was a special quotient expression with constant denominator $t$ and dynamic changing nominator (see Figure 4); and c) that operator $-\partial/\partial t$ casting on sinusoidally changing variable $\Omega_0$ was equivalent to operator cos casting on uniform changing radian $\alpha$ in expressing the same differential operation. The three significances agreed with each other.

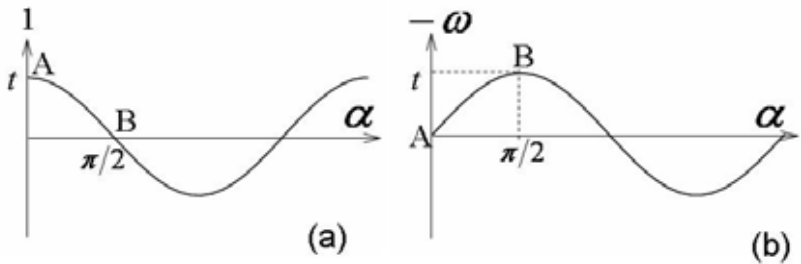

Figure 4. Sinusoidal change of the time component of wave function $\Omega_0$ in (a) cosine composition and (b) sine composition from points A to B where $t$ was a constant of full time dimension in $-\partial\Omega_0/\partial t = (\cos\alpha - i\sin\alpha)$ under Cartesian Coordinates, ignoring the constant amplitude of the complex wave function for brevity here.

## 3. Manifestations of circular motion

Electronic transformation and planetary revolution represented two extremes of circular motion. An electron revolved around the nucleus through changing matter state instead of changing position whereas a planet revolved around the sun through kinematic movement without changing state in itself. Because these two ideal extremes were orthogonal in their nature of circular motion, they can be distinguished by a hypothetical angle $\delta$ in their wave functions. When $\delta$ was equal to $\pi/2$, the orbital motion referred to the pathway of orthogonal state induction; and when $\delta$ was zero, the orbital motion declined to a kinematic circular track. We ordinarily call the former an electronic orbital and the latter a planetary orbit. As shown in Figure 5, these two kinds of ideal circular motions were in two perpendicular planes with a right angle of $\delta$. They had a differential relationship or had $\pi/2$ phase difference in their wave functions in the calculus spacetime.

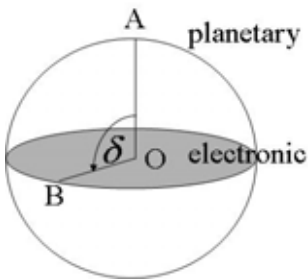

Figure 5. The orthogonal relation of an ideal electronic orbital (shaded circle) and an ideal planetary orbit (blank circle) within two perpendicular planes.

Most of real world orbital motions can be treated as circular motions whose orbits were within the tilted flat planes between the two perpendicular ones, having $\delta$ values in the range of $(0, \pi/2)$. For example, during a year, a longan tree in front of my house experiences four seasons. The plant completes a seasonal cycle similar to a circular motion by changing matter state in response to the weather changes as well as waving with the wind. During the same period, a flock of geese migrate from Canada to the United States in the fall and return to the north in the spring when the weather is getting warmer. The animals are able to lessen the effect of harsh temperature change upon themselves through the strategy of migrations. While the plant has to withstand harsh weather changes by increasing its physical adaptability, the animals choose kinematic movement instead of physical adaptation. The strategies of the plant and the animals in the four seasonal cycling were analogous to the circular motion of electrons and planets, respectively. However, both plants and animals have physical adaptation and kinematic movements in the view scope from electrons to planets, their wave functions must lie in the tilted planes of $0<\delta<\pi/2$.

For real entities whose circular orbit was within a tilted plane ($0<\delta<\pi/2$,), its circular orbit may manifest as an ellipse on the equatorial plane ($\delta=0$) alternatively. Here the equatorial plane of the sun refers to the orbital plane of the nine major planets including the earth. The equivalence of circular orbit on the tilted plane and elliptical orbit on the equatorial plane can be illustrated by the geometrical property of an ellipse as shown in Figure 6. In the first case, a circular orbit with a radius r remained its circular shape (shaded) but its orbital plane rotated a certain angle $\delta$ away from the equatorial plane whereas in the second, an orbit kept in the equatorial plane but deformed into an ellipse (left) whose focus shifted from central point O to $O_1$. When the vertical projection of the circular orbital onto the equatorial plane in the first case was in congruence with the ellipse in the second, we had a relationship concerning the eccentricity of the ellipse:

$$\cos\theta = \sin\delta \tag{17}$$

where $\theta$ was the angle between $OO_1$ and $O_1C$ in right triangle $OO_1C$ where the distance of the focus from the geometric center of the ellipse ($OO_1 = c$) served as a side, semi-minor axis ($OC = b$) served as another side, and semi-major axis ($O_1C = a$) served as the hypotenuse. As $\delta$ angle increased, $\theta$ decreased, and its corresponding ellipse increased its eccentricity (c/a). If $\delta$ increased up to the maximum of right angle rendering the oval became a line with two foci at both ends, then the kinematic movement became matter state transformation as in the case of electronic orbital. The relationship is a well-known geometrical property of an ellipse, but it has a physical implication that goes unnoticed.

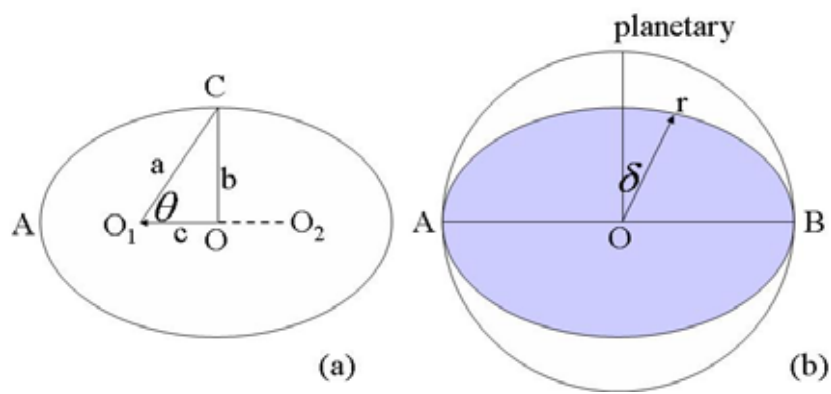

Figure 6 Schematic illustration of the equivalence of (a) an elliptical orbit on the equatorial plane with (b) a circle on a plane tilted $\delta$ angle from the equatorial plane.

The equivalence of a tilted circular motion and an elliptical motion in the equatorial plane reflected the space and time symmetry of nature that has been expressed differentially by equation (14). If we regarded the focus shift $OO_1$ in the equatorial plane as in space, then the tilting departure of the shaded plane from the equatorial plane was in time direction, perpendicular to space. Equation (17) can be interpreted as a calculus relationship in a way similar to equations (14) and (15) that a movement in space was equivalent to a movement in time. The movements in space and in time were only a mathematical trade-off but had different physical manifestations. We believed that electrons traveled in time primarily via state transformation whereas planets took the route of space movements along elliptical orbit during the early genesis. Thus, both electronic orbital and planetary orbit were governed by a unified physical principle and as such Kepler's laws that govern the latter can be logically extended to the former.

## 4. Kepler's law in the atomic spacetime

Kepler's first law states that a planetary orbit is an ellipse with the sun at its focus. What is an ellipse? An ellipse is the projection of a circle onto a plane tilted at a certain angle $\delta$ as was discussed previously. Mathematically, an ellipse is defined as the locus of points, the sum of whose distances from two fixed points, $O_1$ and $O_2$, known as the foci, is a constant (Figure 6a). We believe that such an elegant mathematical property must have its physical significance in the atomic spacetime. To expound this, we need to clarify the physical meaning of the two foci. If one of them $O_1$ is the sun, then what is the other focus $O_2$? Let's go back to electronic orbitals for finding the answer. Under the atomic spacetime, if the movement of $OO_1$ corresponded to the departure of the orbital plane from the equatorial plane at an angle of $\delta$, then the movement of $OO_2$ must be equivalent to the rotation of the orbital plane at the opposite direction in symmetry to $\delta$. Since within helium shell two 1s electrons transformed at opposite directions, they were symmetric about space and time, i.e. space component of one electron was equivalent to time component of the other, and time component of one electron matched space component of the other. Thus, the focus $O_2$ was the location of the sun if we switch space and time components of the solar system. In a helium atom, one electron regarded the location of the nucleus at point $O_1$ while the other saw the location of the nucleus at point $O_2$. But the nucleus was at a fix position, the difference being the electronic orbitals relative to them. Appling the property of the ellipse to the electronic orbitals, we drew a conclusion that the sum of time components of the two electrons was a constant, i.e. time conserved in helium sphere. In symmetry to this, space conserved too. However, the sum must be understood as the addition of two orthogonal quantities that observed Pythagorean theorem as was illustrated by Figure 3:

$$(C_1 \cos\alpha)^2 + (-C_1 \omega \sin\alpha)^2 = C_1^2 \tag{18}$$

$$(C_2 \cos\beta)^2 + |(C_2 r \sin\beta)^2| = C_2^2 \tag{19}$$

where the hypotenuse corresponded to $O_1O_2$ of an ellipse at its maximum eccentricity, which naturally degenerated into a line. Equation (18) indicated electronic time component conservation whereas Equation (19) referred to electronic space component conservation, but they were indeed two aspects of the same process. Equation (16) synchronized the space and time components of an electron in motion, which stated that both radian angles $\alpha$ and $\beta$ were complementary at any moment. Hence Kepler's first law for planetary orbits took the form of Pythagorean theorem for electronic orbitals.

Kepler's second law is closely associated with the first one. To illustrate its analogy in electronic orbital, we need to clarify the relationship between angular velocity $\omega$ and orbital radius $r$. Figure 7 was a combination of both diagrams in Figure 3. It can be seen that orbital radius $r$ was a vector rotating around the origin O while angular velocity $\omega$ was a vector tangential to arc ACB at point C, both orthogonal with their product $v = \omega r$. As point C moved along the arc, angular velocity changed direction smoothly as if rotating around a fix point in Figure 3(a). For a planetary orbit, Kepler's second law states that the line connecting the planet and the sun sweeps equal area within equal time interval. If we interpreted that area swept by that connecting line within a unit time interval in an elliptical orbit as a correspondence of velocity $v$ in the atomic spacetime, then $v$ had a constant magnitude by Kepler's second law disregarding its vector direction. Since electronic transformation is an electromagnetic induction, Kepler's second law dictated constant speed of the electromagnetic wave, which is a well known fact in physics.

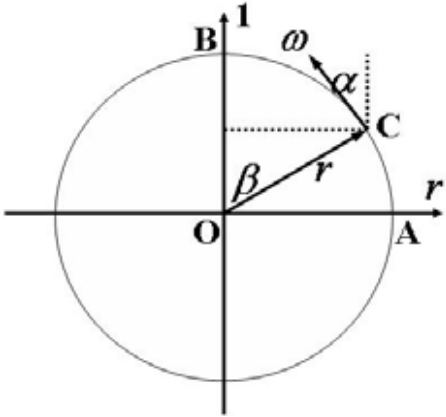

Figure 7. The orthogonal relationship between angular velocity and orbital radius in the atomic spacetime.

For a planetary orbit, Kepler's first law clarifies the elliptical orbit and the second law further specifies speed of movement. Both laws combined to describe an elliptical motion in a certain way. For an electronic orbital, Kepler's first law governed the relation at the dimensional level of $-\partial \Omega_0/\partial t$ and $\int \Omega_1 dl$ quantities while the second law governed the relation at the dimensional level of $-\partial \alpha/\partial t$ and $\int d\beta$ radian operations. When a circular motion was specified at two consecutive orthogonal levels, its orbital was well defined. This was the physical significance of Kepler's two laws from dimensional perspective.

**5. Probability density in Euclidean space**

We have established a coherent architecture of the atomic spacetime in helium. Readers naturally wish to know how to relate quantities in the atomic spacetime to variables in Cartesian coordinates or other Euclidean coordinates that we are familiar with. The answer is not so straightforward as the common translation between Cartesian coordinates and spherical polar coordinates because it is about relating two systems that are rooted in fundamentally different concepts. Nonetheless, the translation between both systems is critical to the success of the atomic spacetime theory because only when it is properly integrated with conventional knowledge can it be recognized and gain its place in science. This section explores a way of geometric translation from abstract electronic transformation of Figure 3 to concrete Euclidean geometry.

As shown in Figure 8, if we interpreted the oscillations of 1s electrons as spherical expansion and contraction in Euclidean geometry, then as electron $\Omega_0$ was undergoing sinusoidal expansion, the other electron $\Omega_2$ experienced harmonic contraction in the meanwhile. At any specific moment, the spherical radius of $\Omega_0$ was determined by OB as point B orbits along a semicircular arc OBA, and the spherical radius of $\Omega_2$ was determined by OC as point C orbits along a semicircular arc ACO, both being synchronized by complementary radian angles $\beta$ in Figure 8(a) and $\alpha$ in Figure 8(b). When point B reached position A, electron $\Omega_0$ attained maximum volume and began to wrap back following the case of $\Omega_2$ in Figure 8(b). On the other hand, when $\Omega_2$ diminished to minimum space at point O, it started to expand following the case of $\Omega_0$ in Figure 8(a). Both electrons inflated and deflated their spatial spheres harmonically with their center at the origin and their radii stretching out or drawing back via the dashed circle, like a crank driving the wheel. However, due to their dimensional difference, both electrons were orthogonal and not at the same physical state during the oscillation.

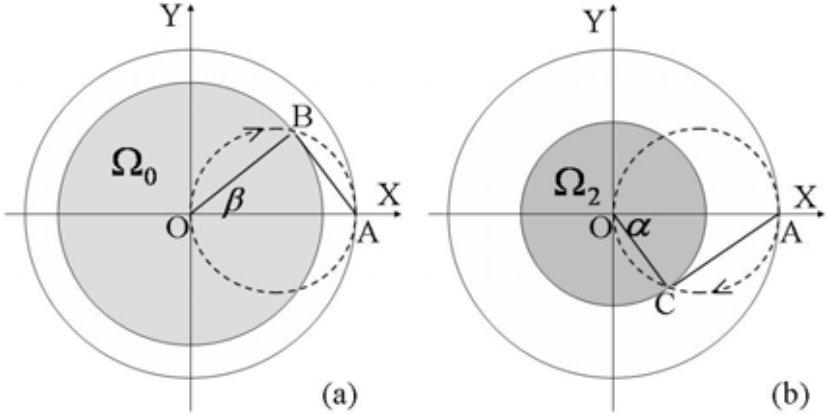

Figure 8. Spherical interpretation of both 1s electrons under Euclidean space where shaded volumes were electron clouds.

Given the foregoing scenario, we may calculate the probability density of electron $\Omega_0$ within the spatial sphere in the case of Figure 8(a). At any specific moment, the spatial volume of the electron was:

$$V = \frac{4}{3}\pi R^3, \tag{20}$$

where the radius $R$ corresponded to chord OB along the dashed circle. If we designated the maximum radius OA as $R_0$, then

$$R = R_0 \cos\beta, \tag{21}$$

where $\beta$ was rotating from $\pi/2$ to 0 uniformly. Suppose electron cloud distributed homogenously within the occupied sphere, then the electronic density must vary with the increment of the spherical volume as follows:

$$dP = -D\frac{4\pi R^2 dR}{V}, \tag{22}$$

where $dP$ meant the differential of electronic density, and $D$ was a constant. Integrating both sides of the equation yields

$$P = \int -D\frac{4\pi R^2}{V} dR. \tag{23}$$

Substituting equation (20) into (23) gives the probability density of the electron cloud at a specific R value as

$$P = -3D \ln(R), \qquad (24)$$

which, upon normalizing $D$ value and ordering $R_0 = 1$, transforms into

$$P = -\ln(\cos\beta), \qquad (25)$$

which indirectly involves two transcendental numbers: $\pi$ in trigonometry and $e$ in natural logarithm.

From an alternative perspective of two-dimensional spacetime, when the space component of the electron was $R_0 \cos\beta$, its time component was $R_0 \sin\beta$ so that

$$dP = -\frac{R_0 \sin\beta}{R_0 \cos\beta} d\beta, \qquad (26)$$

which also leads to equation (25) upon integration. Figure 9 shows the logarithmic function with radius ratio $R/R_0$ in the range of [0,1]. The graph indicates that electronic density was zero at the outer boundary of $R=R_0$ and became infinitely large at the origin of $R=0$. By statistics, this probability density is the chance of finding the electron at R distance from the nucleus through certain physical conversion of the continuous variable to the discrete variable. Although the mathematical result looks simple, exact and elegant, this tentative deduction should be further verified theoretically and experimentally.

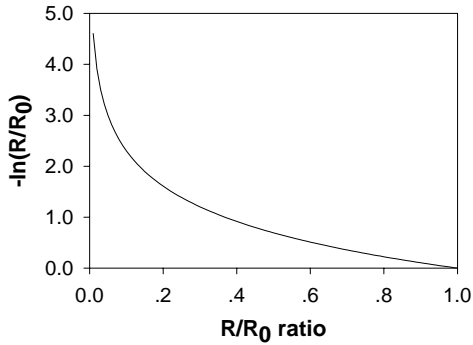

Figure 9. Natural logarithmic function $-\ln(R/R_0)$ as the probability density of an electron at R distance from the nucleus where $R_0$ denotes the maximum radius.

## 6. Central force

Force is of paramount important in physics. For a planetary orbit, Newton deduced his famous gravitational law as follows:

$$F = G\frac{m_1 m_2}{r^2}, \qquad (27)$$

where $m_1$ and $m_2$ are the masses of the two entities, $r$ is the distance between them, $G$ is a proportionality constant, and $F$ is the gravitational force on either object. For electrostatic interaction, Coulomb's law states:

$$F = k\frac{q_1 q_2}{r^2}, \qquad (28)$$

where $q_1$ and $q_2$ are the charges of the two particles and $k$ is a proportionality constant. For centuries, physicists observing the similarity of gravitational law and Coulomb's law have tried to generalize them into a unified theory to cover other interactions. This approach turns out to be nearsighted. After all, Coulomb's law was obtained experimentally rather than theoretically. In addition to experimental evidences, the expression of a formula as complex as Coulomb's law must be a natural result of certain mathematical deduction. If theoretical derivation of its expression has not been made satisfactorily, the law is not completely established and should be regarded as an empirical formula only. In the context of electronic wave function in the atomic spacetime, we shall explore a probable theoretical background of Coulomb's law as follows.

From equation (10), we acquire the first characteristic root to duality equation as

$$\Omega_0 = C_1 C_2 \cos\alpha \cos\beta, \tag{29}$$

where $\alpha$ and $\beta$ were complementary so that by multiple angle formula

$$\Omega_0 = \frac{C_1 C_2}{2} \sin 2\beta, \tag{30}$$

where $\beta$ can be derived from integral operation of $\int 1/r \, dl$ in equation (9). Geometrically, this radian angle was a central angle subtended by a certain arc length $L$ of the radius $r$ of electronic revolution cycle (Figure 10) and can be expressed as

$$\beta = \frac{L}{r}. \tag{31}$$

Substituting the $\beta$ value into equation (30) produces

$$\Omega_0 = \frac{C_1 C_2}{2} \sin \frac{2L}{r}. \tag{32}$$

Since $\Omega_0$ was an inherent characteristic of electronic orbital, it was best interpreted as the energy potential of the electron within helium sphere so that central force that the electron incurred at any moment can be derived from

$$F = -\frac{\partial \Omega_0}{\partial r}, \tag{33}$$

whence

$$F = L\frac{C_1 C_2}{r^2} \cos \frac{2L}{r}, \tag{34}$$

or in terms of infinitive progression

$$F = L\frac{C_1 C_2}{r^2}(1 - \frac{2^2 L^2}{2! r^2} + \frac{2^4 L^4}{4! r^4} - \frac{2^6 L^6}{6! r^6} + ...). \tag{35}$$

Because the infinite series of terms converge, each subsequent term is small relative to its preceding term. If we retained only the first term of the series and neglected the others, then the result would be

$$F \approx L\frac{C_1 C_2}{r^2}, \tag{36}$$

which conforms to equations (27) and (28). Constants $C_1$ and $C_2$ were inherent electronic dimensions in space and time respectively, and the force was the interaction between two 1s electrons. However, if this deduction of force proves to be true and fundamental, then other kinds of interactions may follow suit. Thus if we interpreted $C_1$ and $C_2$ as two orthogonal quantities in general like positive and negative charges, then the force would be electrostatic force. Because orbital radius *r* in the atomic spacetime corresponded to the distance between the orbiting body and the elliptical focus in Euclidean space, our derivation furnished the theoretical basis for Coulomb's law. If so, this indicated that Coulomb's law was not an exact law, but an approximation of equation (35). There have been some experimental reports of the inexactness of Coulomb's electrostatic law, but without theoretical guidance they were generally considered to be within measurement errors. Theoretical attempts for modifying the inverse square law in relativistic or quantum context were normally cumbersome and farther from the truth and hence have never been accepted either.

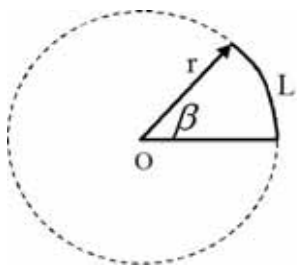

Figure 10. Geometrical relation of $\beta$ as a radian angle and its corresponding arc length *L* in a circle.

Nowadays college physical experiments usually include the measurements of electrostatic forces for verifying Coulomb's law. Students are taught to use the inverse square relationship for curve regression on the data of forces versus distances. When they do not get a satisfactory goodness of fit, they normally blame experimental conditions. To think of the trigonometric term in equation (34) is out of the question. It is believed that a carefully designed experiment should be able to confirm our prediction and estimate the radian angle involved in the expression.

Because electronic motion and planetary movement were manifestations of the same physical principle in different directions, the expression of central force (34) can be applied to gravitational force as well. However, the difference between equations (35) and (36) might be so trivially small for gravitational force that current experimental methods and observations are difficult to discern. While the modification of Newton's gravitational law and Coulomb's law stands to be verified experimentally, we are happy to see that the law of force is a natural outcome of mathematical derivation from primary wave function beyond dirty data analysis.

## 7. Summary

This article has provided an excellent example on the unification of physical entities. Even though planetary orbits and electronic orbitals were remarkably different, they obeyed the same physical principle in essence. While planets orbit around the sun through kinematic movement, electrons looped around the nucleus via matter state transformation. While the planets and orbits belong to different concepts, the electrons and electronic orbitals were indistinguishable. Kepler's first law for planetary orbits took the form of Pythagorean theorem for electrons in the atomic spacetime; and Kepler's second law described the constant velocity of electromagnetic wave. This equivalence demonstrated that the atomic spacetime theory has grasped the fundamentals of nature rather than its diverse superficial manifestations. People have been searching for a grand unification theory in earnest for many centuries, but the Holy Grail of science lies on the more fundamental level than Euclidean space with Newtonian time. One can never manufacture a large ship on board a small boat. It was only on the atomic spacetime platform that various physical entities could be compared and forces and wave functions could be unified in a coherent manner.